\documentclass{article}

\usepackage{geometry}
\usepackage{itm_symbol}
\usepackage{import}
\usepackage{color}
\usepackage{url}
\usepackage{todonotes}
\usepackage{amsmath,amsfonts}
\usepackage{tabularx}
\usepackage{booktabs}
\usepackage[font=small,labelfont=bf]{caption}

\newcommand{\T}{^\mathrm{T}}
\newcommand{\argmin}[1]{\text{arg} \min_{#1}}
\newcommand{\sgn}{\mathrm{sgn}}
\newcommand{\setR}{{\Bbb{R}}}

\newcommand{\meh}[1]{$\,$\mbox{#1}}
\title{Identification of Friction Models for MPC-based Control of a PowerCube Serial Robot}

\author{\underline{J\"org Fehr}$^\ast$, Arnim Kargl$^{\ast}$ and Hannes Eschmann$^{\ast}$}

\begin{document}
\maketitle	
\noindent$^\ast$Institute of Engineering and Computational Mechanics, University of Stuttgart, Germany \\\\
%
% Abstract
%
\noindent\textbf{Abstract}\\
For model-based control, an accurate and in its complexity suitable representation of the real system is a decisive prerequisite for high and robust control quality. In a structured step-by-step procedure, a model predictive control (MPC) scheme for a Schunk PowerCube robot is derived. Neweul-M$^2$ provides the necessary nonlinear model in symbolical and numerical form. To handle the
heavy online computational burden involved with the derived nonlinear model, a linear time-varying MPC scheme is developed based on linearizing the nonlinear system concerning the desired trajectory and the a priori known corresponding feed-forward controller. To improve the identification of the nonlinear friction models of the joints, a nonlinear regression method and the Sparse Identification of Nonlinear Dynamics (SINDy) are compared with each other concerning robustness, online adaptivity, and necessary preprocessing of the input data. Everything is implemented on a slim, low-cost control system with a standard laptop PC.
%
% Introduction
%
\section{Introduction}
The use of robots to increase work performance or human-machine interaction for rehabilitation are topics of high topicality.
For the performance of robotic manipulators, modeling, control, and sensing play an essential role.
Modeling plays an important role in the development process as well as for controlling the robot, e.\,g., for model-based control concepts like model predictive control (MPC). 
Model-based control schemes offer many advantages in comparison to individual joint control. The benefits are: 
(i) operation as a centralized control scheme -- the highly nonlinear behavior of such a system is considered;
(ii) intuitive parameter tuning -- an elaborate and time-consuming parameter tuning necessary for PI or PID-based individual joint control is unnecessary;
(iii) constraint considerations - actuator and state limitations are already considered in the control design.

In \cite{FehrEtAl20} a model predictive controller (MPC) was derived and implemented for a modular 6-axis Schunk PowerCube robot, see Fig.~\ref{img:Picture_Experimental_Setup}. 
The general MPC algorithm, see, e.\,g.,~\cite{RawlingsMayne09,Maciejowski02} is the following: (i) obtain a measurement/estimate of the state; (ii) obtain an optimal input sequence by solving online an open-loop optimal control problem (OCP) over a finite horizon subject to system dynamics and constraints; (iii) apply the first input of the optimal input sequence to the plant/robot; (iv) continue with the first step.
Challenges for the application are the highly nonlinear system dynamics as well as the limited computation capacity. Everything should run on a slim low-cost setup, i.\,e., a standard laptop PC.

\begin{figure}[htpb]
	\begin{minipage}[t]{0.475\textwidth}
		\centering
		\def\svgwidth{0.5\columnwidth}
		\includegraphics[scale=1.0]{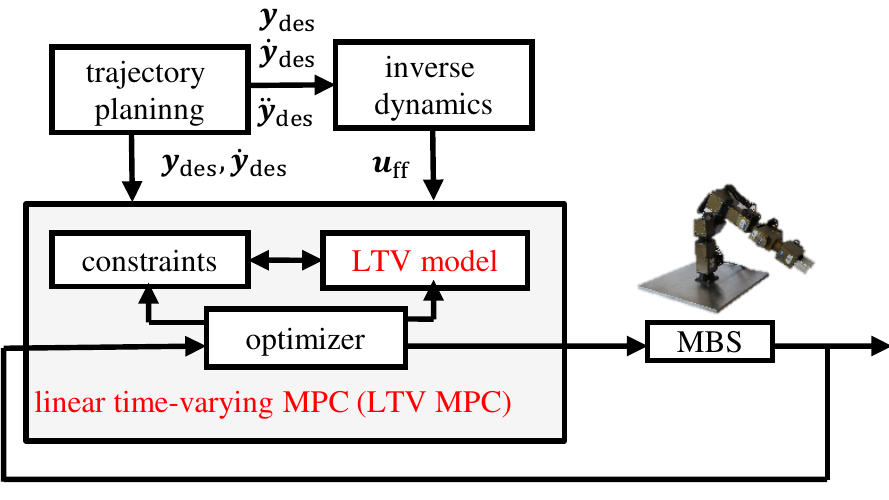}
		\caption{Topologic structure of the two-step control loop with the feed-forward part
			$\ub_{\mathrm{ff}}$ and the LTV-MPC part.}
		\label{fig:Overall_Control_Structure}
	\end{minipage}
	\begin{minipage}[t]{0.05\textwidth}
		\centering
		\hfill
	\end{minipage}
	\begin{minipage}[t]{0.475\textwidth}
		\centering
		\includegraphics[scale=1.0]{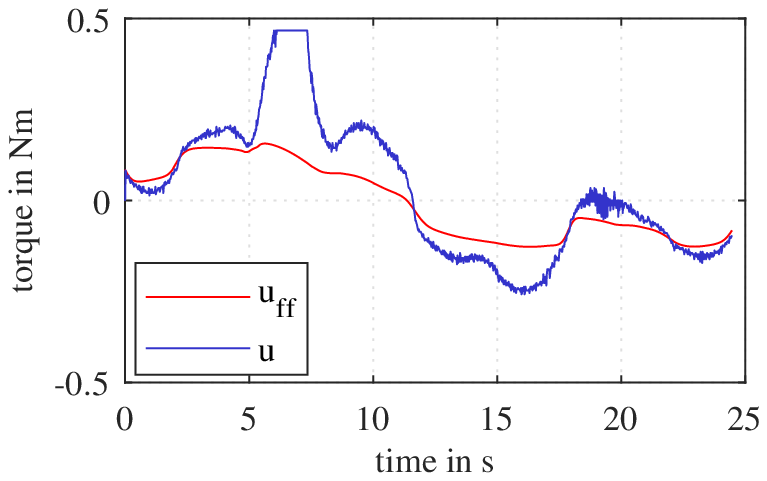}
		\caption{Feed forward torque of the old robot model compared to the MPC output torque
			applied at joint $\beta$ during a trajectory with motion reversal.
			\label{fig:feedForwardComparison_original}}
		%		\includegraphics{fig/mpcVelocityError_english}
		%		\caption{Performance of the non-optimized MPC controller for joint C during a typical
		%			trajectory used to perform system identification of the joint friction.						
		%			\label{fig:mpcVelocityError}}
	\end{minipage}
\end{figure}

These challenges are met by a linear time-variant two-step approach, see Fig.~\ref{fig:Overall_Control_Structure}. In a first step the nonlinear system dynamics are approximated with a linear time-varying model~\cite{SchnelleEberhard14} around the desired trajectory --using an  inverse dynamic approach / computed torque approach~\cite{GrotjahnHeimann02}. In the second step sophisticated techniques for linear MPC are exploited, i.\,e., the open-source quadratic programming solver qpOASES~\cite{FerreauEtAl14} is used.

The proposed LTV MPC control of the robotic system is able to perform complex trajectories, i.\,e., motion reversal and zero crossing ~\cite{FehrEtAl20}
\footnote{The complete motion of the manipulator 
	can be seen in the deposited video~\url{https://www.itm.uni-stuttgart.de/en/research/vision-based-control-of-a-powercube-robot/}.}.
Nevertheless, the noticeable difference between calculated feed forward and MPC output, see Fig.~\ref{fig:feedForwardComparison_original} along the trajectory implies model inaccuracies. 
%The non-optimal tracking performance of the manipulator shows that the model of the robot still has some weaknesses.
These findings imply that the friction model of the robot's joints still has some weaknesses which limits the performance of the overall control system.
Therefore, more effort is needed to identify the friction properties of the joints. 
Prior to this work, a classical system identification method, i.\,e., data pre-processing in combination  with nonlinear regression, was used to identify the friction properties of a single disassembled module.

Therefore, aim of this work is to evaluate how the large amount of data from different sources, i.\,e., Artifical Intelligence will facilitate the complex and challenging modeling and system identification of an assembled system -- e.\,g., can the large amount of data be used to identify the effects in the assembled state and/or difference between individual products due to production tolerances.

More specifically, we want to answer the question of whether the Sparse Identification of Nonlinear Dynamics (SINDy) method ~\cite{BruntonProctorKutz16,KaiserKutzBrunton18} improve the friction identification.
In contrast to black-box AI methods, like Neural Networks or Gaussian Processes -- which try to approximate the data by adjusting some weights of a topological system, the 
SINDy method tries to identify the governing equations from data. It approximates an unknown function $\fb$ with a library $\Thetab(\Xb)$ of $r$ potential (nonlinear) terms. 
The SINDy approach is a parametric approach that, compared to NNs, works without massive amounts of data.
The approach allows for on-the-fly model adaptation due to its low computational complexity.

Let us recap the overall goal: "Improve the model-based control performance of a robotic manipulator by improving friction identification using the SINDy approach to identify the governing equation from data."
The main features are: (i) the robustness of the approach and (ii) only the friction characteristic is identified -- other well-identified or known terms of the system are incorporated as prior knowledge.

%The following part of the paper is structured in the following way.
In the next section, we describe the model and the methodology in more detail: (i) the robot which serves as an example; (ii) the process control framework; (iii) the derivation of the equation of motion of the rigid multibody system with Neweul-M\textsuperscript{2}~\cite{KurzEtAl10} (iv) the existing friction model, (v) the SINDy concept and (vi) the recording of measurement data.
In Section Friction Identification the results of the approach are presented.
Finally, in the Conclusion the overall control performance of the system with the improved friction model is discussed.

%
%Thanks to the vast improvement of computational power during the last decades, the implementation is realized with a standard laptop PC as low-cost environment. \\
%Nevertheless, the significant intervention of the feedback part of the control  Those weaknesses are mainly founded in modeling joint friction. 
%In Fig.~\ref{fig:mpcVelocityError} it is possible to identify the tracking error of an individual joint while tracking the desired trajectory with MPC. \\

%In this paper, a data-based identification method will be combined with model-based control approaches to learn friction models of the robotic joints adaptively. More specifically, we want to answer the question can the friction identification be improved by the Sparse Identification of Nonlinear Dynamics (SINDy) method ~\cite{BruntonProctorKutz16,KaiserKutzBrunton18} in comparison to a classical nonlinear regression model in combination with data pre-processing.

%%%%%%%%%%%%%%%%%%%%%%%%%%%%%%%%%%%%%%%%%%%%%%%%%%%%%%%%%%%%
% Model and Methodology
%%%%%%%%%%%%%%%%%%%%%%%%%%%%%%%%%%%%%%%%%%%%%%%%%%%%%%%%%%%
\section{Model and Methodology}
\begin{figure}[b]
	\begin{minipage}[b!]{1.0\textwidth}
		\begin{minipage}[t]{0.4\textwidth}
			\centering
			\def\svgwidth{0.5\columnwidth}
			\includegraphics[scale=1.0]{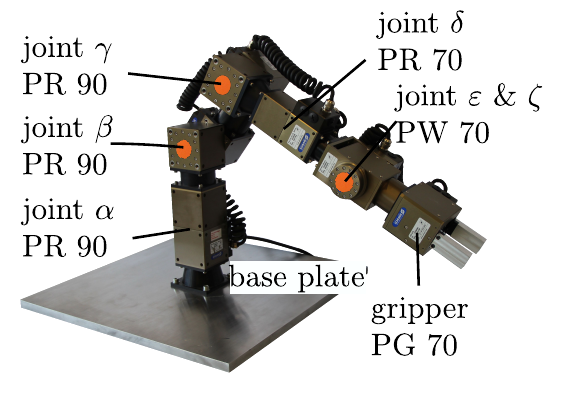}
			\captionof{figure}{Experimental setup and configuration of the six degrees of freedom Schunk modular
				manipulator. The control is based on Real-Time Windows Target and Simulink Coder. 
				\label{img:Picture_Experimental_Setup}}
		\end{minipage}
		\hfill
		\begin{minipage}[b]{0.6\textwidth} 
			\captionof{table}{Technical properties of the Schunk modules.\label{tab:SchunkModules}}
			{\begin{tabular}{p{7em}p{8.5em}p{7.135em}}
					\toprule
					module name & nom. torque/force & max. velocity \\
					\midrule
					PR 90 (rotary) joint $\alpha$ \& $\beta$  \& $\gamma$ & $44.8 \meh{Nm}$ & $\omega_{\text{max}}=25 \meh{rpm}$ \\
					PR 70 (rotary) joint $\delta$ & $10.0 \meh{Nm}$ & $\omega_{\text{max}}=25 \meh{rpm}$ \\
					PW 70 (pan-tilt) joint $\varepsilon$ \& $\zeta$ & $12.0 \meh{Nm}$ \& $2.0 \meh{Nm}$ & $\omega_{\text{max}}=25 \meh{rpm}$ \\
					PG 70 (gripper) & $200 \meh{N}$ & $v_{\text{max}}=82 \meh{mm/s}$ \\
					\bottomrule
			\end{tabular}}
		\end{minipage}
	\end{minipage}
\end{figure}

The performance of the SINDy approach is evaluated on 
a modular 6-axis \textit{Schunk PowerCube Robot}, see Fig.~\ref{img:Picture_Experimental_Setup}. For the considered modular robot a \textit{Process Control Framework} in Matlab Simulink and a \textit{Simulation Model} in Neweul-M$^2$ is available.
We focus on the friction identification of the first three joints of the robot, which are each a rotary unit PR 90 with a Harmonic Drive gear to see the influence of assembly and production differences of the same product. Therefore measurement data is gathered during experiments followed by offline identification of the friction characteristics with nonlinear regression and the proposed \textit{SINDy Framework}, relying on \textit{Sparse Linear Regression}. In the following, we explain the single building blocks for model, \textit{Experimental Measurements} and mathematical methodology.

\paragraph{Schunk PowerCube Robot}
\label{par:SchunkPowerCubeRobot}
$\:$ \\
\\ 
The properties of the various links are described in Tab.~\ref{tab:SchunkModules}.
For the design of the robot a "divide et impera" approach is used. For the first three links the same rotary module (PR~90) is used. The fourth link, is the smaller version(PR~70) within the model series. The fifth and six links of the robot consist of a pan-tilt unit in combination with an anthropomorphic gripper with a spherical wrist, is constructed from the single modules. 
The robotic manipulator is designed in such a way that the arm has an analytically calculable inverse kinematics.

\begin{figure}[htpb]
	\begin{minipage}[t]{0.8\textwidth}
		\centering
		\def\svgwidth{0.5\columnwidth}
		\includegraphics[scale=1.0]{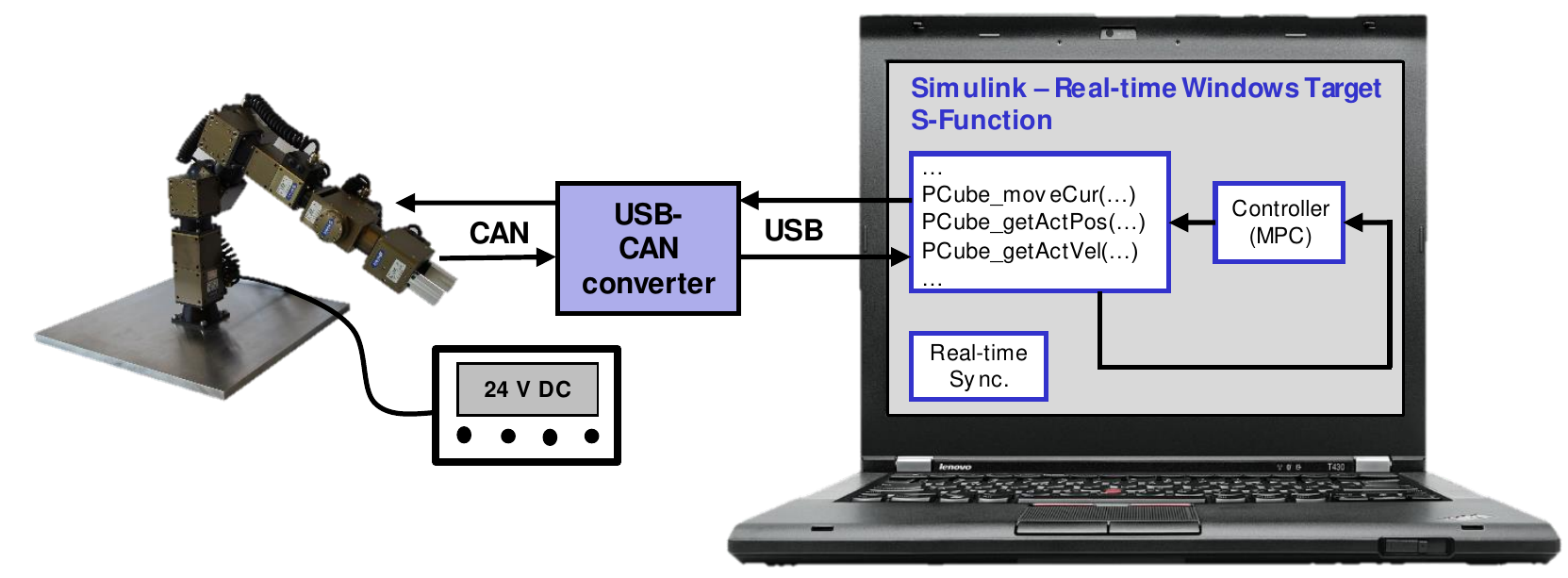}
		\caption{Slim process control framework of the robotic manipulator }
		\label{fig:Process_Control_Framework}
	\end{minipage}
	
\end{figure}

Each rotary module, consists of a brushless DC-motor which drives a Harmonic Drive gear, which provides torque at each degree of freedom~(DOF) based on the defined motor current. The control and power electronics are integrated, and an incremental encoder is used for position and speed evaluation. Furthermore, a brake is incorporated in case of shutdown or power failure.
\paragraph{Process Control Framework}
\label{par:ProcessControlFramework}
$\:$ \\
\\
The process control framework is depicted in Fig.~\ref{fig:Process_Control_Framework} , consists of a Microsoft Windows laptop PC with Matlab R2014b and Simulink, including the additional toolboxes Real-Time Windows Target and Simulink Coder. The laptop PC is equipped with an Intel Core i5-3210M CPU (2x2.5\,GHz, 8\,GB DDR3). The communication between the Simulink model and the hardware is based on an USB-CAN bus interface, which is embedded into Simulink via S-Functions and a communication library from Schunk. A sampling rate of 50 Hz, corresponding to a sampling interval of 20\,ms, is used. An external power supply with constant voltage of 24\,V in combination with the integrated control and power electronics of modules  ensures the necessary torques at the links. 

Real-Time Windows Target~\cite{MatlabRTWT14} realizes a real-time engine for Simulink models on a Microsoft Windows PC and offers the capability to run hardware-in-the-loop simulations in real-time. It is a lean solution for rapid prototyping and provides an environment in which a single computer can be used as a host and target computer. Consequently, real-time simulations are executed in Simulink without an external target machine.

\paragraph{Simulation}
\label{par:Simulation}
$\:$ \\
The robotic manipulator is modeled as rigid multibody system with joint friction. In a first step Neweul-M$^2$ \cite{KurzEtAl10} aids calculating a rigid body model \textit{without friction} with the advantage of generating equations of motion in symbolic and numerical form. A natural choice for generalized coordinates $\yb$ are the joint coordinates $\yb = [\beta, \gamma]\T$. Therefore the resulting equation of motion (\textit{without friction}) in minimal form can be denoted as
\begin{equation}
\label{eq:mbsWithoutFriction}
\Mb(\yb) \ybpp + \kb(\yb, \ybp) = \qbs(\yb, \ybp) + \Bb \ub
\end{equation}
with the positive definite mass matrix $\Mb$, the vector of generalized Coriolis, centrifugal and gyroscopic forces $\kb$ and the vector of generalized forces without friction $\qbs$. The system input is the vector $\ub = [T_2, T_3]\T$ which consists of the applied motor torque at each joint. Those motor torques are scaled accordingly with gear ratios included in $\Bb$. A linear model $\ub = K_{\mathrm{M}} \ib$ describing the relation between motor torque and motor current is assumed. The motor constant $K_{\mathrm{M}}$ has been identified manually at the hardware. 
System parameters, such as masses, inertias, link lengths and gear ratios are taken from CAD data and datasheets, supplied by the manufacturers. \\
In a second step friction torques $\taub(\ybp)$ are included into the model without friction, resulting in the vector of generalized forces
\begin{equation}
\qb(\yb, \ybp) = \qbs(\yb, \ybp) - \taub(\ybp). 
\end{equation}
For a more intuitive understanding $\taub$ is subtracted from $\qbs$, since positive friction works against the current movement. The equation of motion \textit{with friction} can be gained by substitution of $\qbs$ with $\qb$ in eq.~\eqref{eq:mbsWithoutFriction}.
Joint friction models of the form 
\begin{equation}
\label{eq:stribeckFriction}
\tau_i(\dot{y}_i) = a_1 \dot{y}_i + a_2 \tanh(a_3 \dot{y}_i)
+ a_4 \exp(- a_5 \vert \dot{y}_i \vert) \tanh(3 a_3 \dot{y}_i),
\end{equation}
derived in \cite{Oberhuber13} are used in a first application which results in the stated mismatch between feed forward and MPC output. This form of friction model with parameter vector $\ab$ includes the share of viscous friction corresponding to $a_1 \dot{y}_i$, the share of Coulomb friction corresponding to $a_2 \tanh(a_3 \dot{y}_i)$ as well as the superelevation of the Stribeck curve which is characterized by $a_4 \exp(- a_5 \vert \dot{y}_i \vert) \tanh(3 a_3 \dot{y}_i)$. The smooth \textit{tanh}-function replaces the \textit{sgn}-function to avoid discontinuities within the model.
\\
Focus lies on identifying more accurate friction models for both joints with either the SINDy method or general nonlinear regression. Therefore data of $\yb$, $\ybp$, $\ybpp$ and $\ub$ has to be collected during experiments or derived after running experiments respectively, such that
\begin{equation}
\taub(\ybp) = \qbs(\yb, \ybp) + \Bb \ub - \Mb(\yb) \ybpp - \kb(\yb, \ybp)
\end{equation}
can be calculated. Stated methods can then be applied. The resulting values for $\taub$ obviously are dependent on all stated variables $\yb$, $\ybp$, $\ybpp$ and $\ub$. The notation $\taub(\ybp)$ is chosen due to the assumption that friction effects only depend on joint velocities.
\\

\begin{figure}[hptb]
	\begin{minipage}[t]{0.475\textwidth}
		\centering
%% Accompanies image file 'SINDyIllustration.pdf' (pdf, eps, ps)
\begingroup%
  \makeatletter%
  \providecommand\color[2][]{%
    \errmessage{(Inkscape) Color is used for the text in Inkscape, but the package 'color.sty' is not loaded}%
    \renewcommand\color[2][]{}%
  }%
  \providecommand\transparent[1]{%
    \errmessage{(Inkscape) Transparency is used (non-zero) for the text in Inkscape, but the package 'transparent.sty' is not loaded}%
    \renewcommand\transparent[1]{}%
  }%
  \providecommand\rotatebox[2]{#2}%
  \newcommand*\fsize{\dimexpr\f@size pt\relax}%
  \newcommand*\lineheight[1]{\fontsize{\fsize}{#1\fsize}\selectfont}%
  \ifx\svgwidth\undefined%
    \setlength{\unitlength}{162.46923455bp}%
    \ifx\svgscale\undefined%
      \relax%
    \else%
      \setlength{\unitlength}{\unitlength * \real{\svgscale}}%
    \fi%
  \else%
    \setlength{\unitlength}{\svgwidth}%
  \fi%
  \global\let\svgwidth\undefined%
  \global\let\svgscale\undefined%
  \makeatother%
  \begin{picture}(1,0.58954375)%
    \lineheight{1}%
    \setlength\tabcolsep{0pt}%
    \put(0,0){\includegraphics[width=\unitlength,page=1]{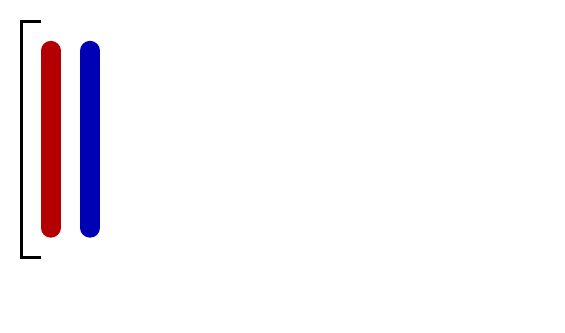}}%
    \put(0.24645227,0.32218727){\color[rgb]{0,0,0}\makebox(0,0)[lt]{\lineheight{1.25}\smash{\begin{tabular}[t]{l}$=$\end{tabular}}}}%
    \put(0,0){\includegraphics[width=\unitlength,page=2]{SINDyIllustration.pdf}}%
    \put(0.0515635,0.03952294){\color[rgb]{0,0,0}\makebox(0,0)[lt]{\lineheight{1.25}\smash{\begin{tabular}[t]{l}$\xbp_1$\end{tabular}}}}%
    \put(0.13879988,0.03952294){\color[rgb]{0,0,0}\makebox(0,0)[lt]{\lineheight{1.25}\smash{\begin{tabular}[t]{l}$\xbp_2$\end{tabular}}}}%
    \put(0.77613436,0.03952294){\color[rgb]{0,0,0}\makebox(0,0)[lt]{\lineheight{1.25}\smash{\begin{tabular}[t]{l}$\xib_1$\end{tabular}}}}%
    \put(0.86337075,0.03952294){\color[rgb]{0,0,0}\makebox(0,0)[lt]{\lineheight{1.25}\smash{\begin{tabular}[t]{l}$\xib_2$\end{tabular}}}}%
    \put(0.40657908,0.03952294){\color[rgb]{0,0,0}\makebox(0,0)[lt]{\lineheight{1.25}\smash{\begin{tabular}[t]{l}$\Thetab(\Xb, \Ub)$\end{tabular}}}}%
  \end{picture}%
\endgroup%

		\caption{Application of the SINDy concept with sparse linear regression to a system with
			two states. Grey entries in $\xib_k$ mark entries which are zero.}
		\label{img:sindyIllustration}
	\end{minipage}
	\begin{minipage}[t]{0.05\textwidth}
		\centering
		\hfill
	\end{minipage}
	\begin{minipage}[t]{0.475\textwidth}
		\centering
%% Accompanies image file 'STLSIllustration.pdf' (pdf, eps, ps)
\begingroup%
  \makeatletter%
  \providecommand\color[2][]{%
    \errmessage{(Inkscape) Color is used for the text in Inkscape, but the package 'color.sty' is not loaded}%
    \renewcommand\color[2][]{}%
  }%
  \providecommand\transparent[1]{%
    \errmessage{(Inkscape) Transparency is used (non-zero) for the text in Inkscape, but the package 'transparent.sty' is not loaded}%
    \renewcommand\transparent[1]{}%
  }%
  \providecommand\rotatebox[2]{#2}%
  \newcommand*\fsize{\dimexpr\f@size pt\relax}%
  \newcommand*\lineheight[1]{\fontsize{\fsize}{#1\fsize}\selectfont}%
  \ifx\svgwidth\undefined%
    \setlength{\unitlength}{227.35017935bp}%
    \ifx\svgscale\undefined%
      \relax%
    \else%
      \setlength{\unitlength}{\unitlength * \real{\svgscale}}%
    \fi%
  \else%
    \setlength{\unitlength}{\svgwidth}%
  \fi%
  \global\let\svgwidth\undefined%
  \global\let\svgscale\undefined%
  \makeatother%
  \begin{picture}(1,0.4337686)%
    \lineheight{1}%
    \setlength\tabcolsep{0pt}%
    \put(0,0){\includegraphics[width=\unitlength,page=1]{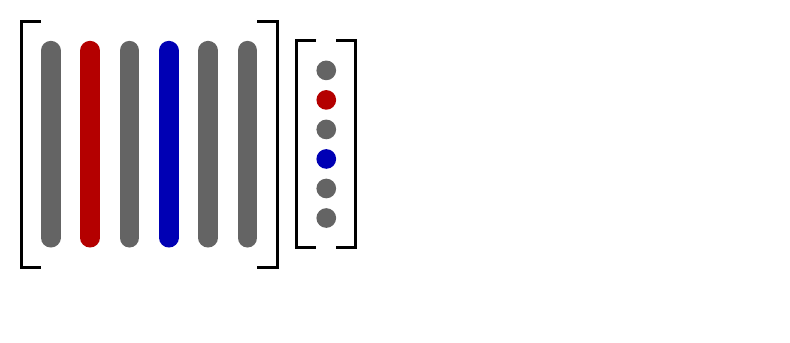}}%
    \put(0.15539826,0.02824393){\color[rgb]{0,0,0}\makebox(0,0)[lt]{\lineheight{1.25}\smash{\begin{tabular}[t]{l}$\Thetab^{(r)}$\end{tabular}}}}%
    \put(0,0){\includegraphics[width=\unitlength,page=2]{STLSIllustration.pdf}}%
    \put(0.3918295,0.02824393){\color[rgb]{0,0,0}\makebox(0,0)[lt]{\lineheight{1.25}\smash{\begin{tabular}[t]{l}$\xib_k^{(r)}$\end{tabular}}}}%
    \put(0.67884144,0.02824393){\color[rgb]{0,0,0}\makebox(0,0)[lt]{\lineheight{1.25}\smash{\begin{tabular}[t]{l}$\Thetab^{(r+1)}$\end{tabular}}}}%
    \put(0.82285233,0.11552127){\color[rgb]{0,0,0}\makebox(0,0)[lt]{\lineheight{1.25}\smash{\begin{tabular}[t]{l}$\xib_k^{(r+1)}$\end{tabular}}}}%
  \end{picture}%
\endgroup%

		\caption{Illustration of the STLS algorithm. Grey entries in $\xib_k^{(r)}$ are set
			to zero. Corresponding columns in $\Thetab^{(r)}$ are removed. $\xib_k^{(r+1)}$ is the new least
			squares solutions with slightly different entries.}
		\label{img:stlsIllustration}
	\end{minipage}
\end{figure}

\paragraph{SINDy Concept}
\label{par:SINDy Concept}
$\:$ \\
The concept of \textbf{S}parse \textbf{I}dentification of \textbf{N}onlinear \textbf{Dy}namics (SINDy) founds in the field of applied mathematics. It represents a modern method to gain nonlinear models based on experiment data, a long known challenge in system theory. The main concept behind SINDy can be described as reducing the nonlinear fit to a collection of function candidates to a (sparse) linear regression, which can then be effectively solved with state of the art algorithms, providing robust and efficient solutions. \\
The SINDy setup consists of a standard representation of a nonlinear system
\begin{equation}
\xbp = \fb(\xb, \ub)
\end{equation}
with state vector $\xb \in \setR^n$, possible input $\ub \in \setR^q$ and the unknown vector field $\fb$. Starting out simple, assume the state $\xb$, and its time derivative $\xbp$ and the system input $\ub$ to be known for $m$ unique time instances. The data can then be rearranged into three matrices $\Xb$, $\Xbp$ and $\Ub$ as follows
\begin{equation}
\Xb =
\begin{bmatrix}
\xb \T (t_1) \\
\xb \T (t_2) \\
\vdots \\
\xb \T (t_m)
\end{bmatrix}
\in \setR^{m \times n}, \quad
\Xbp =
\begin{bmatrix}
\xbp \T (t_1) \\
\xbp \T (t_2) \\
\vdots \\
\xbp \T (t_m)
\end{bmatrix}
\in \setR^{m \times n}, \quad
\Ub =
\begin{bmatrix}
\ub \T (t_1) \\
\ub \T (t_2) \\
\vdots \\
\ub \T (t_m)
\end{bmatrix}
\in \setR^{m \times q}.
\end{equation}
Goal is to approximate the vector field $\fb$ by a library of $r$ candidate functions $\Theta(\Xb, \Ub) \in \setR^{m \times r}$ which are weighed by coefficients $\Xib = [\xib_1, \xib_2, \cdots, \xib_r] \in \setR^{n \times r}$ as
\begin{equation}
\label{eq:sindyLinearRegression}
\xbp = \fb(\xb, \ub) \approx \Thetab(\Xb, \Ub) \Xib,
\end{equation}
where the columns of the library of candidate functions can contain all kinds of imaginable nonlinear terms. Those can be polynomials in $\xb$, $\ub$ or combinations of both or other nonlinear terms like trigonometric functions. Filling this library with appropriate terms requires intuition on how the solution could look and is considered one critical point concerning the success of the SINDy method.
\\
The solution to the linear regression problem introduced in eq.~\eqref{eq:sindyLinearRegression} may be calculated with aid of various solvers, e.\,g., a standard least squares solver
\begin{equation}
\label{eq:standardLeastSquares}
\xib_k = \argmin{\wb}
\big\Vert \Xbp_k - \Thetab(\Xb, \Ub) \wb \big\Vert_2^2.
\end{equation}
Here $\xib_k$ denotes the $k$-th column of $\Xib$ while $\Xbp_k$ denotes the $k$-th column of $\Xbp$. \\
Doing so does result in a reasonable model but at the same time leads to very complex and detailed models with many active function terms. Key feature of the SINDy concept is solving the linear regression problem eq.~\eqref{eq:sindyLinearRegression} promoting sparsity within the solution vectors $\xib_k$, which corresponds to the fact that most systems of interest can be described by only a few active terms in $\fb$. Therefore solution techniques for sparse linear regression are applied to eq.~\eqref{eq:sindyLinearRegression}. A schematic overview and example for a system with two states is given with Fig.~\ref{img:sindyIllustration}.
\\
The overall concept appears to be very flexible regarding to all kinds of dynamical systems. There exists a wide range of extensions which , e.\,g., allow SINDy to perform in discrete time or even to identify PDEs \cite{BruntonProctorKutz16,RudyEtAl17}.
\\\\
The SINDy concept, in general, brings some advantages and some disadvantages compared to other machine learning concepts, e.\,g., an artificial neural network (NN). Such a NN may be used to describe the unknown vector field $\fb$, which also gives good results. One general disadvantage of a NN is the  training process, which requires  large amounts of measureed data to obtain an suitable approximation for $\fb$. SINDy on the other hand can work with one single short experimental trajectory as in Fig.~\ref{fig:feedForwardComparison_original}. The SINDy method results in a set of nonlinear differential equations which may even be physically interpretable where a NN just results in some non-interpretable matrix layers. In addition to that, the SINDy results, being differential equations, can be evaluated in their whole domain and possibly even be extrapolated. This is a major advantage over neural networks, which often provide unsatisfying approximation results outside of the domain of the training data. Training data of course, has to cover all important system characteristics for both approaches. Otherwise, some system aspects and properties will be left out in the identified system model.\\
A crucial drawback of the SINDy concept lies in the function library $\Thetab$ and in the fact that only linear combinations of these functions describe the resulting dynamics $\fb$ of the system.
One critical disadvantage of the SINDy concept lies in its function library $\Thetab$ and that only linear combinations of those functions can be described in the resulting $\fb$.
Looking at eq.~\eqref{eq:stribeckFriction}, one cannot directly identify a nested term like $a_2 \tanh(a_3 \dot{y}_i)$ with SINDy.
Having a look back at eq.~\eqref{eq:stribeckFriction} one cannot directly identify a term like $a_2 \tanh(a_3 \dot{y}_i)$ with SINDy.  The problem is that the parameter $a_3$ is embedded inside a nonlinear term. We will later avoid this by including multiple terms $\tanh(a_3 \dot{y}_i)$ with different fix parameters $a_3$ into the library $\Thetab$. \\
The main advantages and problems of the SINDy concept have been briefly touched upon here. In \cite{KaiserKutzBrunton18}, however, a much more detailed comparison of SINDy with artificial neural networks when used in model predictive control can be found.\\
%
%\begin{minipage}[t]{0.475\textwidth}
%%	\includegraphics{fig/mpcVelocityError_english}
%%    \caption{Performance of the non-optimized MPC controller for joint \beta during a typical
%%    trajectory used to perform system identification of the joint friction.						
%%    \label{fig:mpcVelocityError}}
%\end{minipage}
%\end{figure}

\paragraph{Sparse Linear Regression}
$\:$ \\
Two algorithms out of the wide variety of sparse linear regression methods were tested and could be applied to the setup. One of them is the so called Least Absolute Shrinkage and Selection Operator (LASSO), which is well known from the field of statistics and often used in machine learning applications. The LASSO can be described as $\ell_1$ regularized version of the standard least squares linear regression setup and can be written as
\begin{equation}
\xib_k = \argmin{\wb}
\big\Vert \Xbp_k - \Thetab(\Xb, \Ub) \wb \big\Vert_2^2
+ \lambda \big\Vert \wb \big\Vert_1.
\end{equation}
The parameter $\lambda \ge 0$ adjusts the influence of the penalty term $\Vert \wb \Vert_1$ and therefore determines how sparse the solution vector $\xib_k$ turns out. With $\lambda = 0$ the solution is equal to the standard least squares problem's solution stated in eq.~\eqref{eq:standardLeastSquares}. \\
Another sparsity promoting linear regression approach is given by the Sequential Thresholded Least Squares algorithm (STLS) introduced in \cite{BruntonProctorKutz16}. The STLS algorithm is an iterative procedure based on standard least squares solutions and can be described as follows.
\begin{itemize}
	\item An initial solution $\xib_k^{(0)}$ is calculated as least squares solution with the
	complete library matrix $\Thetab^{(0)} = \Thetab(\Xb, \Ub)$;
	\item In each iteration the regression setup reduces by
	\begin{itemize}
		\item setting entries in $\xib_k^{(r)}$ with absolute value less than a threshold parameter
		$\lambda$ to zero;
		\item deleting the corresponding columns in $\Thetab^{(r)}$, which would then be multiplied
		by zero, which leads to $\Thetab^{(r+1)}$;
	\end{itemize}
	Solving the reduced least squares problem with $\Thetab^{(r+1)}$ leads to a new solution
	vector $\xib_k^{(r+1)}$ which includes the remaining entries of $\xib_k^{(r)}$.
	\item Iteration ends if no more entries in $\xib_k^{(r)}$ fulfill the threshold condition.
\end{itemize}
Again, the parameter $\lambda > 0$ determines the sparsity of the solution. One iteration step of the STLS algorithm is illustrated in Fig.~\ref{img:stlsIllustration}. \\
The critical point with both algorithms is choosing an appropriate value for the hyper parameter $\lambda$.
In practice models for a broad range of parameters $\lambda$ are calculated which allows finding a good compromise between sparsity of the resulting model and model error along a pareto front. The LASSO combined with $k$-fold cross validation is able to find this compromise by its own. With the STLS approach as it is described above one has to choose a model by hand.
More sparse solutions on the one hand allow for some form of physical interpretation of the result but on the other hand may show a greater model error compared to other more flexible nonlinear regression techniques. \\\\
The whole SINDy setup was implemented in Matlab~\cite{MathWorks19}. The STLS algorithm was implemented manually while a LASSO implementation already exists within the Statistics and Machine Learning Toolbox.

\begin{figure}[hptb]
	\begin{minipage}[t]{0.475\textwidth}
		\centering
		\includegraphics[scale=1.0]{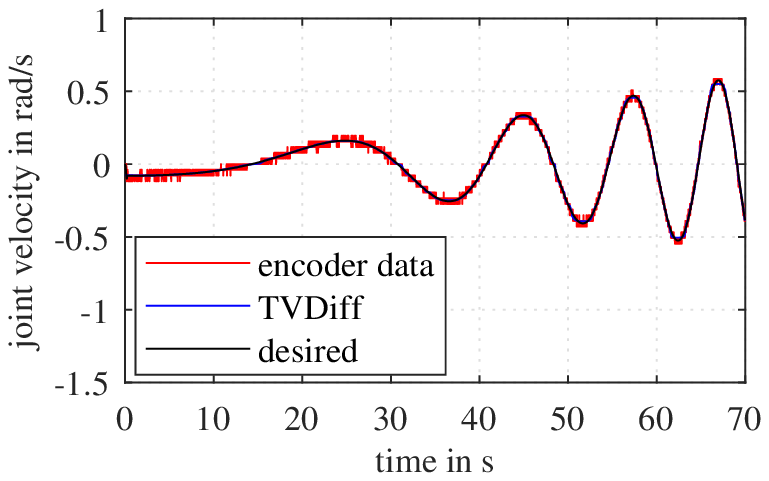}
		\caption{Measured encoder velocity compared with the filtered velocity from position measurements
			via TVDiff and the desired velocity.}
		\label{fig:compareVelocityMeasures}
	\end{minipage}
	\begin{minipage}[t]{0.05\textwidth}
		\centering
		\hfill
	\end{minipage}
	\begin{minipage}[t]{0.475\textwidth}
		\centering
		\includegraphics[scale=1.0]{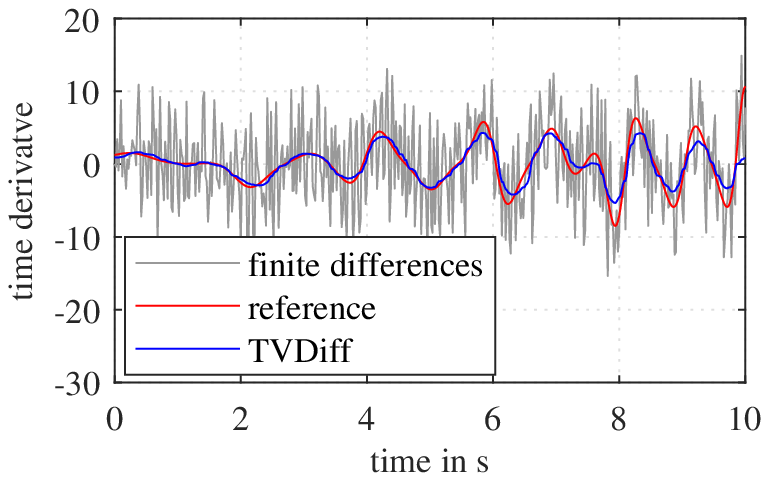}
		\caption{Synthesized example to demonstrate TVDiff advantages over finite differences
			with comparison to the analytically calculated reference signal.}
		\label{fig:diffDemoDerivative}
	\end{minipage}
\end{figure}

\paragraph{Experimental Measurements}
$\:$ \\
Before taking measurements, a suitable class of trajectories has to be defined used for friction identification. We chose sine-shaped trajectories with variable frequency within this study since they can be constructed relatively easily. Being at least two times continuous differentiable sine trajectories brings smooth acceleration and deceleration, which is crucial for not exceeding joint limitations. Furthermore, differentiation and integration can be done analytically. The sine part of the trajectory can be described as
\begin{equation}
y(t) = \hat{a} \: \sin(\omega(t) t)
\end{equation}
with a constant amplitude $\hat{a}$.
Polynomial acceleration and deceleration phases around the sine trajectory part are needed since the robot starts and ends in a static pose where velocities and accelerations must be zero. Part of such a trajectory can be seen in Fig.~\ref{fig:compareVelocityMeasures}. \\
The figure also points out a difficulty with measurement data. The robot's joint positions are measured by absolute angle encoders within each joint, which results in a non-smooth velocity measurement with coarse resolution. Therefore a suitable differentiation method is needed to calculate joint velocities and accelerations from joint position data. In \cite{KaiserKutzBrunton18} the use of Total Variation Regularization Differentiation (TVDiff) is recommended. The TVDiff algorithm is introduced in \cite{Chartrand11}. The paper and Matlab implementations of the algorithm for one- and two-dimensional data, can be found on the author's web page.
The idea of TVDiff originates in the Tikhonov regularization where the energy of a signal is minimized according to an energy functional without influencing the signal in a way a low pass filter would do. Variational methods are quite popular in the field of imaging science for efficiently denoising images. TVDiff results for trajectory measurement data are presented in Fig.~\ref{fig:compareVelocityMeasures}. Figure.~\ref{fig:diffDemoDerivative} demonstrates the advantages of TVDiff over finite differences with a therefore synthesized signal with added white noise.
With higher frequencies a slight low pass effect is visible but the results are sufficient for the application with SINDy.

\section{Friction Identification}
\begin{figure}[b]
	\begin{minipage}[t]{0.475\textwidth}
		\centering
		\includegraphics[scale=1.0]{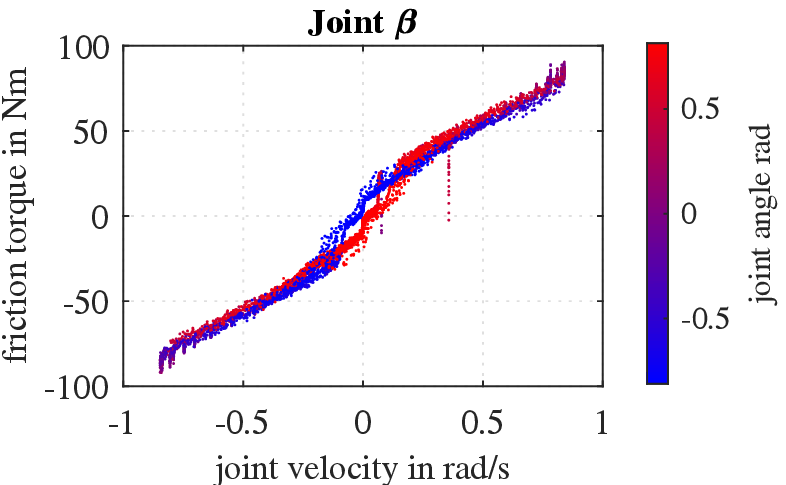}
		\caption{Friction characteristic for joint B measured with a sine trajectory. Only joint B
			was moved for these measurements.}
		\label{fig:betaT15angle}
	\end{minipage}
	\begin{minipage}[t]{0.05\textwidth}
		\centering
		\hfill
	\end{minipage}
	\begin{minipage}[t]{0.475\textwidth}
		\centering
		\includegraphics[scale=1.0]{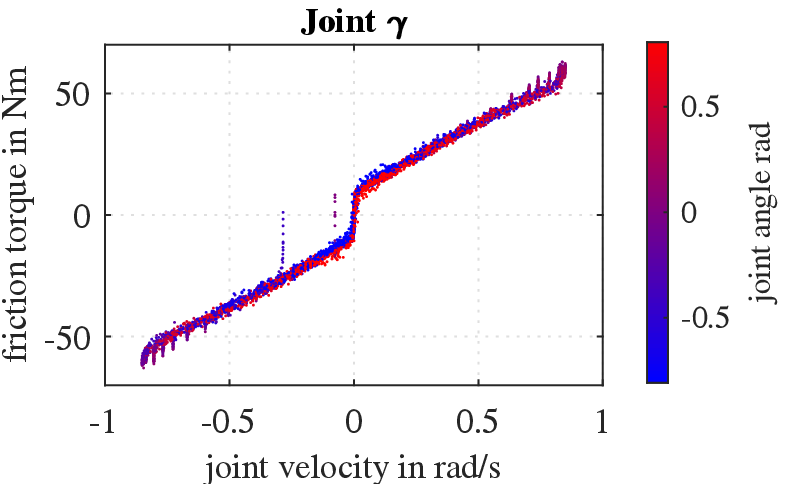}
		\caption{Friction characteristic for joint C measured with a sine trajectory. Joint B and
			joint C were moved equally in opposite directions to compensate gravity.}
		\label{fig:gammaT15angle}
	\end{minipage}
\end{figure}

With previously described methods almost all infrastructure needed to identify friction models is present. Friction characteristics for robot joints B and C which correspond to the joint coordinates $\beta$ and $\gamma$ are displayed in Fig.~\ref{fig:betaT15angle} and Fig.~\ref{fig:gammaT15angle}.

\begin{figure}[ptb]
	\centering
	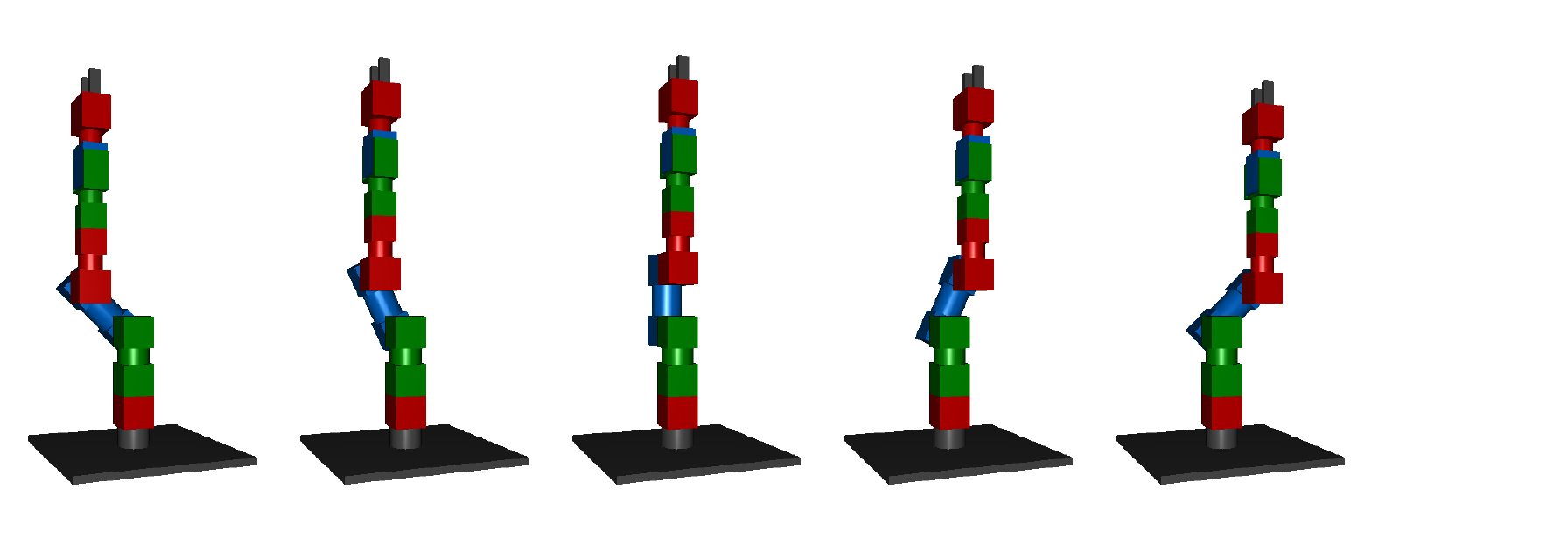
	\caption{Periodic trajectory for friction identification in joint C. Shown is half a period
		starting at the left turning point (a), going through the robot's zero position (c) and ending
		at the right turning point (e). Frames (b) and (d) display intermediate points.}
	\label{img:trajectory15}
\end{figure}

\begin{figure}[tb]
	\begin{minipage}[t]{0.475\textwidth}
		\centering
		\includegraphics[scale=1.0]{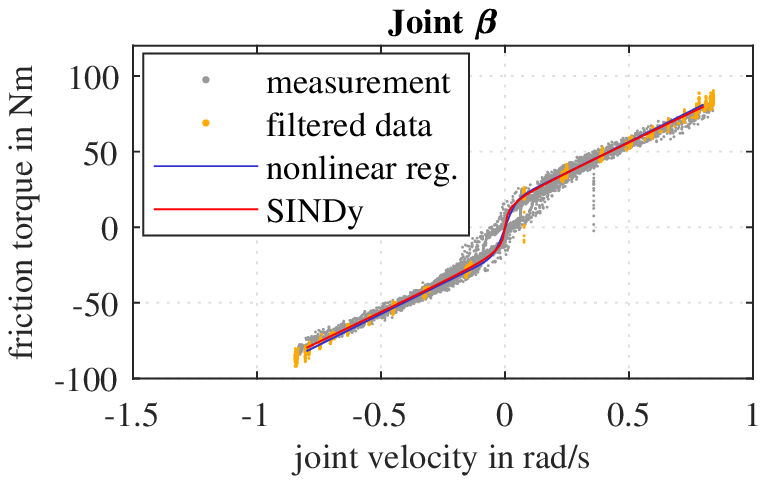}
		\caption{Fitted friction models for joint B after preprocessing measurement data.}
		\label{fig:betaT15nlreg}
	\end{minipage}
	\begin{minipage}[t]{0.05\textwidth}
		\hfill
	\end{minipage}
	\begin{minipage}[t]{0.475\textwidth}
		\centering
		\includegraphics[scale=1.0]{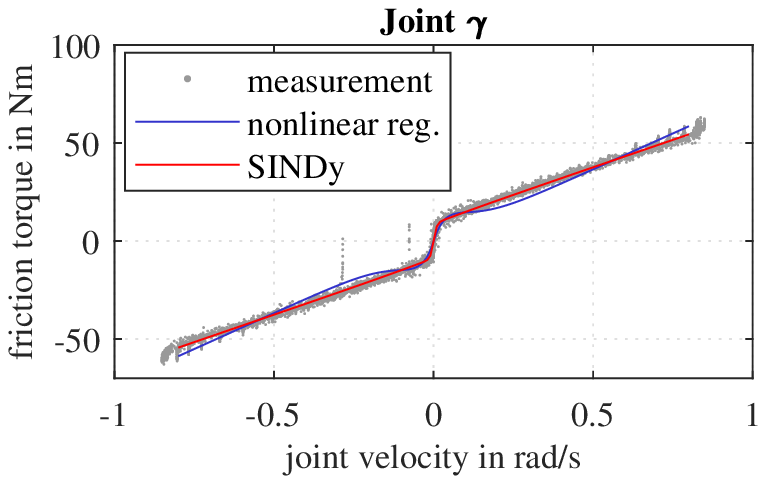}
		\caption{Fitted friction models for joint C. No preprocessing is needed.}
		\label{fig:gammaT15nlreg}
	\end{minipage}
\end{figure}

\begin{figure}[hptb]
	\begin{minipage}[t]{0.475\textwidth}
		\centering
		\includegraphics[scale=1.0]{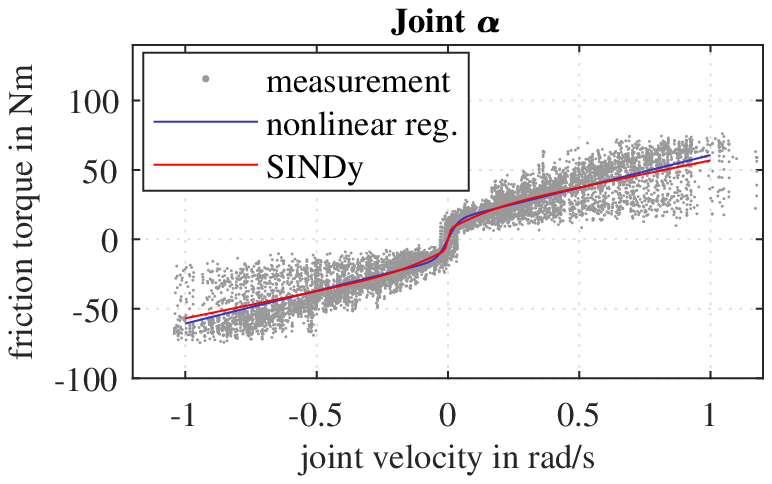}
		\caption{Fitted friction models for joint A. No preprocessing is needed.}
		\label{fig:alphaT16nlreg}
	\end{minipage}
	\begin{minipage}[t]{0.05\textwidth}
		\centering
		\hfill
	\end{minipage}
	\begin{minipage}[t]{0.475\textwidth}
		\centering
		\includegraphics[scale=1.0]{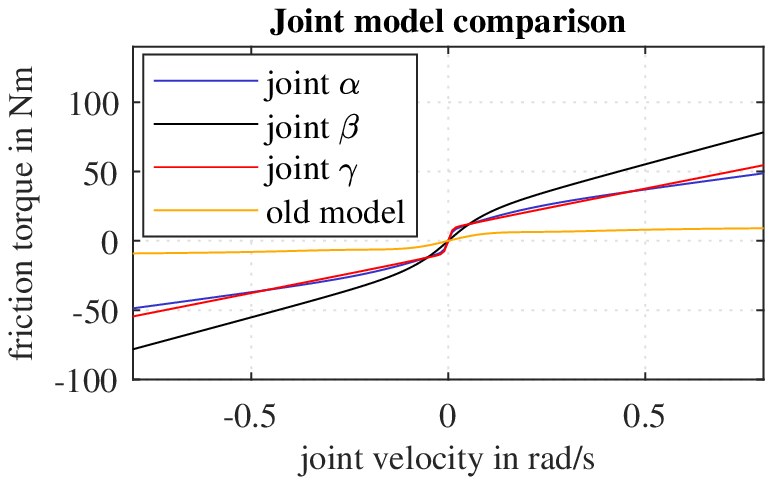}
		\caption{Comparison of the three new joint friction models with the old model from
			\cite{Oberhuber13}. Visualized are the identified SINDy models.}
		\label{fig:jointComparison1SE}
	\end{minipage}
\end{figure}

The resulting friction characteristics of both joints, if moved individually, will be dependent on the joint angle. These effects arise from model inaccuracies within the rigid multibody dynamics, e.\,g., incorrect inertia values or not modeling cables etc. The influence of gravity at non-zero joint angles makes the resulting friction characteristics dependent on the joint angle.
This is against the assumption of friction only depending on joint velocity. Therefore for joint B the gained data is preprocessed by selecting data points with $\vert \beta \vert < 0.1$ where the influence of gravity is negligible. The filtered data is plotted in Fig.~\ref{fig:betaT15nlreg}. Since many data points remain unused, a different strategy is chosen for joint C. Since it is the upper body of two joints, the influence of gravity can be easily compensated by moving both joints in opposite directions, keeping the upper part of the robot pointing straight upwards. Figure~\ref{img:trajectory15} illustrates half of a period of the particular periodic trajectory. Identification with the gravity compensated trajectory leads to the friction characteristics for joint C shown in Fig.\ref{fig:gammaT15angle}.
As stated earlier the SINDy results are compared to a model calculated by nonlinear regression. Eq.~\ref{eq:stribeckFriction} therefore functions as template for the nonlinear regression with its parameters $a_1$ to $a_5$. The nonlinear regression algorithm used is the \texttt{nlinfit} Matlab function which is included in the Statistics and Machine Learning Toolbox as well. \\
The application of the SINDy concept still requires us to specify a library of function terms. Over time and within many experiments the library
\begin{equation}
\label{eq:sindyFrictionLibrary}
\Thetab(\ybp) = 
\begin{bmatrix}
\bm{1} & \ybp & \sgn(\ybp) & \tanh(5 \ybp) & \tanh(10 \ybp) & \tanh(20 \ybp)
& \tanh(100 \ybp)
\end{bmatrix}
\end{equation}
lead to promising results. The library contains functional terms for describing friction characteristics and at the same time, leaves enough options for the sparse linear regression to cancel out function terms.
Promoting sparsity without leaving enough options for the algorithm would not lead to the expected results. 
A finer sampling of the terms $\tanh(a \ybp)$ was tested but results in high correlations between these functions, ultimately leading to difficulties in sparse linear regression. The library does not contain terms describing the superelevation of the Stribeck curve. This will only be a minor disadvantage since such a superelevation is not noticeable within the measurement data. \\
Applying SINDy and nonlinear regression to the measurement data together with the earlier discussed preprocessing leads to the the friction models shown in Fig.~\ref{fig:betaT15nlreg} and Fig.~\ref{fig:gammaT15nlreg}. 
Both regression concepts deliver quite similar results. SINDy outperforms nonlinear regression in the case of joint C, having the ability to choose freely from its function library while the nonlinear regression always has to fit to its function template. While nonlinear regression for the data of joint B would not work properly without preprocessing, the SINDy method would still lead to reasonable results. Although the preprocessed data gives a better friction model and shows that SINDy can even work with a percentage of the available data, making it robust to the amount of available measurement data. \\
An additional friction identification was performed for link A (corresponding to $\alpha$) for better comparison of the joint friction characteristics and models. Therefore a linear quadratic regulator was implemented, since there doesn't exist an MPC scheme for the rigid body model with $\yb = \alpha$. The resulting friction models for link A are displayed in Fig.~\ref{fig:alphaT16nlreg}. Again, nonlinear regression as well as the SINDy method perform well, although the measurement data is spread quite heavily. Preprocessing the data is not necessary for either method. \\
Comparing the newly identified joint friction models to the previously used model from \cite{Oberhuber13} points out two major aspects. At first, friction models differ from joint to joint, although being the same kind of link module (PR90). Secondly, whether the joint friction is identified for an isolated joint or at an assembled robot makes a big difference. Therefore the two newly identified joint friction models differ heavily from the model identified in \cite{Oberhuber13}. While in \cite{Oberhuber13} single isolated joint modules were researched, the additional weight of our robot above joints B and C amplifies friction at the joints' axles. Fig.~\ref{fig:jointComparison1SE} shows the comparison of the different models. Compared are the SINDy results identified above.
%\cite{Oberhuber13}
%\todo{{Oberhuber13} add to caption in final version} 
\begin{figure}[hptb]
	\begin{minipage}[t]{0.475\textwidth}
		\centering
		\includegraphics[scale=1.0]{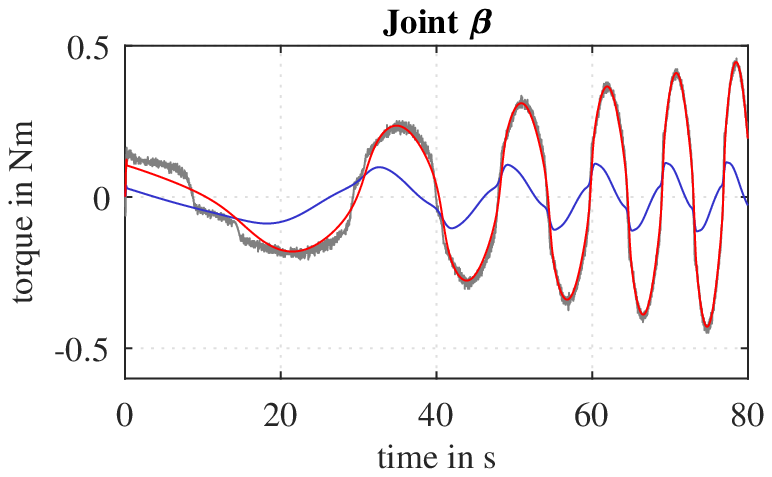}
		\caption{Feed forward torqes of the old and new model compared to the MPC output torque
			applied at joint $\beta$.}
		\label{fig:feedForwardComparisonBeta}
	\end{minipage}
	\begin{minipage}[t]{0.05\textwidth}
		\centering
		\hfill
	\end{minipage}
	\begin{minipage}[t]{0.475\textwidth}
		\centering
		\includegraphics[scale=1.0]{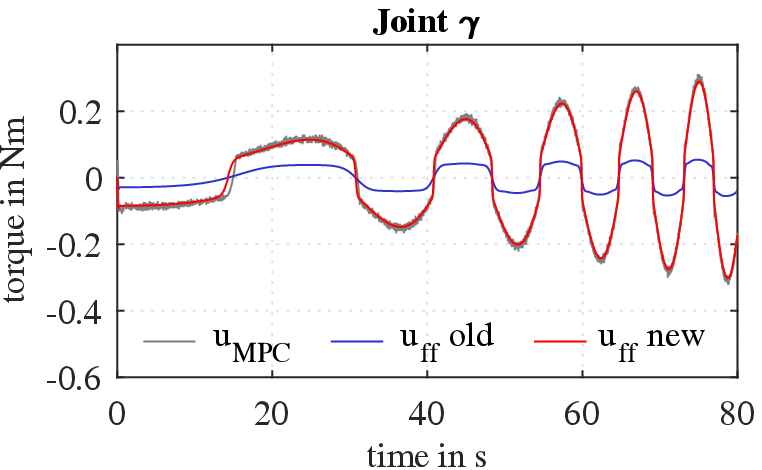}
		\caption{Feed forward torqes of the old and new model compared to the MPC output torque
			applied at joint $\gamma$.}
		\label{fig:feedForwardComparisonGamma}
	\end{minipage}
\end{figure}

\section{Conclusion}
With the methods and results developed throughout this paper, we can finally find the following conclusion.
Both, System Identification for Nonlinear Dynamics (SINDy) and nonlinear regression were applied to identify friction models for the relevant joints B and C of the PowerCube robot. Additionally, a friction model for joint A was identified for use with a linear quadratic regulator.
The vast improvement of the feed-forward calculations together with the newly identified joint friction models imply a general improvement of the robot model. Although the feed-forward torque almost matches the controller output, slight differences still remain that result from model errors within the rigid body model without friction. 

However, better trajectory tracking performance, critical for practical usage of the robotic manipulator, could not be improved with the current setup. A first reason lies in using a linearizing MPC approach that does not adequately represent nonlinear system dynamics. Second, the setup underlies limitations in hardware and software. On the hardware side, communication between the robot and the centralized controller in Simulink is limited in speed since the Schunk PowerCube Robot is designed for decentralized joint control schemes such as PID. On the software side, the MPC online optimization cannot be evaluated arbitrarily fast. Both limitations contrast the need for trajectories and linearized dynamics with high resolution in time whenever using LTV MPC. \\
Further research out of the scope of this publication has shown improvements in trajectory tracking performance with the newly identified friction models when using a truly nonlinear MPC scheme.

When determining the friction,  the research has been shown that the friction determination for isolated joint connection modules is not sufficient. Within the robot assembly, the axes of the joints have to bear an additional load, which leads to higher friction.\\
The SINDy method and its sparse regression algorithms make for a robust and fast method for identifying nonlinear system dynamics. Nonlinear regression in comparison leads to very similar results but needs more computation time due to its nonlinear optimization, making it less capable for real-time applications. \\
All in all, SINDy's potential to identify parts of unknown system dynamics was successfully presented.

\subsection{Acknowledgements}
Funded by Deutsche Forschungsgemeinschaft (DFG, German Research Foundation) under Germany's Excellence Strategy - EXC 2075 - 390740016. We acknowledge the support by the Stuttgart Center for Simulation Science (SimTech).
We thank Patrick Schmid and Georg Schneider for the helpful discussions regarding MPC and the experimental setup.

%
% Bibliography
%
%\bibliographystyle{itm_paper_engl}
%\bibliography{ITM_Literatur,Eigene_Literatur}

\end{document}